\pdfoutput=1

\documentclass[aps,prd,twocolumn,showpacs,superscriptaddress,groupedaddress, twoside,notitlepage,preprintnumbers,showpacs,showkeys,superscriptaddress,byrevtex,floatfix,amsmath,amssymb]
{revtex4-1}
\usepackage{array,dsfont,graphicx,mathtools,placeins,verbatim,tensor,youngtab}
\usepackage[bookmarksnumbered,bookmarksopen=true,bookmarksopenlevel=1,pdfstartview=FitH]{hyperref}
\usepackage[all]{hypcap}
\newcolumntype{M}[1]{>{$}{#1}<{$}}

\newcommand{\sst}[1]{{\scriptscriptstyle #1}}

\newcommand{\rep}[1]{\ensuremath{\mathbf{#1}}}
\allowdisplaybreaks

\def\0{{\sst{(0)}}}
\def\1{{\sst{(1)}}}
\def\2{{\sst{(2)}}}
\def\3{{\sst{(3)}}}
\def\4{{\sst{(4)}}}
\def\5{{\sst{(5)}}}
\def\6{{\sst{(6)}}}
\def\7{{\sst{(7)}}}

\def\oi{{\sst{(01)}}}
\def\io{{\sst{(10)}}}
\def\ii{{\sst{(11)}}}

\def\oj{{\sst{(02)}}}
\def\jo{{\sst{(20)}}}
\def\jj{{\sst{(22)}}}

\def\ij{{\sst{(12)}}}
\def\ji{{\sst{(21)}}}
\def\jj{{\sst{(22)}}}

\newcommand{\be}{\begin{equation}}
\newcommand{\ee}{\end{equation}}
\def\ba{\begin{array}}
\def\ea{\end{array}}

\newcommand{\bea}{\begin{eqnarray}}
\newcommand{\eea}{\end{eqnarray}}

\DeclareMathOperator{\USp}{USp}
\DeclareMathOperator{\SL}{SL}

\DeclareMathOperator{\Sp}{Sp}

\DeclareMathOperator{\Spin}{Spin}
\newcommand{\tG}{\tilde{G}}

\newcommand{\tA}{\tilde{A}}

\newcommand{\R}{\mathds{R}}

\newcommand{\Z}{\mathds{Z}}

\newcommand{\N}{\mathcal{N}}

\begin{document}

\title{$D=6$, $\N=(2,0)$ and $\N=(4,0)$ theories}  
\author{L. Borsten}
\email[]{leron@stp.dias.ie}
\affiliation{School of Theoretical Physics, Dublin Institute for Advanced Studies,
10 Burlington Road, Dublin 4, Ireland}

\date{\today}

\begin{abstract}
 Using a convolutive field-theoretic product, it is shown here that the ``square'' of an Abelian  $D=6$, $\N=(2,0)$  theory yields  the free $D=6$, $\N=(4,0)$  theory constructed by Hull, together with  its generalised (super)gauge transformations. This  offers a new perspective on the $(4, 0)$ theory and chiral theories  of conformal gravity more generally, while at the same time extending the domain of the ``gravity = gauge $\times$ gauge''   paradigm. 

\end{abstract}


\preprint{DIAS-STP-17-07}

\maketitle

\section{Introduction}

It was at one time thought that non-trivial conformal quantum field theories exist in at most $D=4$ spacetime dimensions. This was somewhat  at odds with Nahm's   classification  of admissible supersymmetries \cite{Nahm:1977tg}, which  includes $D=6$  superconformal algebras.  Indeed, a remarkable prediction of M-theory \cite{Witten:1995ex, Strominger:1995ac, Maldacena:1997re},  anticipated in \cite{Gunaydin:1984wc, Blencowe:1987bn},  is the existence of non-trivial $D=6$ quantum  field theories   with $\N=(2,0)$ supersymmetry and   $\text{OSp}^\star(8|4)$ superconformal symmetry, contradicting the received wisdom of the time while  placing another feather  in Nahm's cap. 
 These ``$(2, 0)$ theories'' are  not only central to our understanding of M-theory; they have fundamental implications for lower-dimensional gauge theories more generally, from   S-duality to the Alday-Gaiotto-Tachikawa (AGT) correspondence \cite{Kapustin:2006pk,Witten:2009at, Alday:2009aq}.

Of course,   the  consistency  of a given superalgebra  does not   imply that a corresponding  non-trivial quantum field theory  necessarily  exists. See for example \cite{Cordova:2016emh}. However, taking confidence from the $(2, 0)$ story it is tempting  to speculate that the  $D=6, \N=(4, 0)$  multiplet with $\text{OSp}^\star(8|8)$ superconformal symmetry,  a longstanding and enticing outpost of   Nahm's taxonomy, should also correspond to a non-trivial quantum  theory. Indeed, drawing on a range of analogies with the (2,0) theories Hull argued  \cite{Hull:2000zn, Hull:2000ih, Hull:2000rr} that a non-trivial  ``$(4, 0)$ theory'' may arise in the large $D=5$ Plank length, $l_5$, limit of M-theory compactified on 6-torus, $T^6$. 
As emphasised by Hull,  the $(4, 0)$ theory would constitute the   maximally symmetric phase of M-theory. Moreover, it contains a self-dual ``gravi-gerbe'' field, suggestive of a $D=6$ chiral  theory of  conformal gravity. Note, a local variational principle, breaking \emph{manifest} covariance,  for the free gravi-gerbe field was recently developed in \cite{Henneaux:2016opm}. Consequently,  just as for the $(2,0)$ theories before it,  establishing its existence would have profound implications for not only  M-theory, but also gravity more broadly understood.  It should be stressed that while there is a large body of strong evidence, originating from string/M-theory, for the (2,0) theories, there are at present no comparable arguments supporting the existence of the (4,0) theory and it remains highly conjectural. For a more nuanced discussion of the various possibilities, and the associated difficulties, the reader is referred to \cite{Hull:2000zn, Hull:2000ih, Hull:2000rr, Schwarz:2000zg, Chiodaroli:2011pp}.

Here we re-examine the free $(4,0)$ theory introduced in \cite{Hull:2000zn, Hull:2000ih, Hull:2000rr} from another, a priori unrelated, but equally provocative,   perspective: ``gravity = gauge $\times$ gauge''.  While on face-value a radical proposal, 
this paradigm has been reinvigorated in recent years by  the remarkable  Bern-Carrasco-Johansson  double-copy procedure \cite{Bern:2008qj, Bern:2010ue, Bern:2010yg}; the scattering amplitudes of (super)gravity are conjectured to be the ``double-copy'' of (super) Yang-Mills amplitudes to all orders in perturbation theory! These fascinating amplitude relations are both computationally expedient and conceptually suggestive, facilitating previously intractable calculations while probing profound  questions regarding  the  deep structure of perturbative quantum gravity \cite{Bern:2012cd, Bern:2014sna}.

In this context $D=5$, $\N=8$ supergravity, the low energy limit  of M-theory  on a 6-torus,  is  the double-copy of $D=5$, $\N=4$ super Yang-Mills theory.   Of course, $D=5$   Yang-Mills theory is non-renormalisable and we expect new physics to enter 	for energies $E\geq 1/g_{YM}^{2}$. For instance, it can be regarded as the low-energy sector  of the world-volume theory of a stack of D$4$-branes in string theory. Taking the strong-coupling  limit the  Yang-Mills theory uplifts to a  $(2, 0)$ theory  compactified on a circle  of radius $R\propto g^{2}_{YM}$, which in this setting constitutes  the low-energy theory arising on a stack of  M5-branes in M-theory. This raises  a challenging question: what happens to the double-copy in this limit? Might we expect some relation of the type $(4, 0)=(2,0)\times (2, 0)$, morally the   M-theory uplift of   gravity = gauge $\times$ gauge?

The $(4, 0)=(2,0)\times (2, 0)$ picture  was proposed in \cite{Chiodaroli:2011pp}, where the  ultra-short $(4, 0)$ supermultiplet of the six-dimensional conformal superalgebra $\text{OSp}^\star(8|8)$ was derived and shown to  consistently factorise, with respect to the R-symmetry algebras $\USp(4)\times\USp(4)\subset\USp(8)$, into the product of two $(2,0)$ tensor multiplets. However, as emphasised in \cite{Chiodaroli:2011pp} the intrinsically non-perturbative nature of the $(2, 0)$ theories makes amplitude relations hard to formulate, although there exist some limited tests \cite{Huang:2010rn, Chiodaroli:2011pp, Czech:2011dk}. Here we avoid this hurdle altogether by appealing to a complementary and independent off-shell field-theoretic realisation of gravity as the ``square of Yang-Mills'' developed in \cite{Borsten:2013bp, Anastasiou:2013hba, Anastasiou:2014qba, Nagy:2014jza, Anastasiou:2015vba, Borsten:2015pla, Cardoso:2016ngt, Cardoso:2016amd, Anastasiou:2016csv, Anastasiou:2017nsz}, which can be used to study the product of two gauge theories without reference to amplitudes, allowing one to derive various properties, such as  curvatures, dynamics, off-shell local symmetries and duality relations, directly. For two gauge potentials belonging to  two distinct Yang-Mills theories, referred to as the   left (no tilde) and  right (tilde) factors, with  arbitrary gauge groups $G$ and $\tilde{G}$, the product is given by  \cite{Anastasiou:2014qba}:
\be\label{def}
A_\mu \circ \tA_{\nu}:= A_{\mu}^{a}  \cdot \Phi_{a\tilde{a}} \cdot \tA_{\nu}^{\tilde{a}},
\ee
 where
$
  [f\cdot g](x)=\int d^Dy f(y)g(x-y).
 $
The bi-adjoint  ``spectator'' scalar field   $\Phi$   allows for arbitrary and independent   $G$ and $\tG$, while   the convolution reflects the fact that the amplitude relations are multiplicative in momentum space. Crucially, together they  ensure that  both the global and local symmetries of the two factors are consistently mapped into those of the corresponding gravitational theory, including general coordinate transformations  \cite{Borsten:2013bp, Anastasiou:2013hba, Anastasiou:2014qba, Nagy:2014jza, Anastasiou:2015vba, Borsten:2015pla}. To linear approximation the equations of motion of the factors then imply those of the gravity theory and classical solutions of the Yang-Mills  factors are  mapped into solutions of their product \cite{Anastasiou:2014qba, Cardoso:2016ngt, Cardoso:2016amd}.  Extending this construction  it is shown here that, by defining a field \eqref{G} and ghost field \eqref{paras} dictionary,  
     the product of two arbitrary  Abelian $(2,0)$ theories generates, with no further input, the free $(4, 0)$ theory first constructed by Hull \cite{Hull:2000zn}.  This  represents a new perspective on the $(4, 0)$ theory that may be exploited to better understand its remarkable,  as yet rather mysterious, properties, while  at the same time extending the rapidly  evolving domain \cite{Anastasiou:2013hba, Anastasiou:2014qba, Nagy:2014jza, Anastasiou:2015vba, Borsten:2015pla, Cardoso:2016ngt, Cardoso:2016amd, Anastasiou:2016csv, Anastasiou:2017nsz, Monteiro:2011pc,Monteiro:2013rya,Fu:2016plh, Carrasco:2012ca, Damgaard:2012fb, Huang:2012wr, Bargheer:2012gv, Johansson:2014zca, Monteiro:2014cda, Chiodaroli:2014xia, Chiodaroli:2015rdg, Chiodaroli:2015wal, Luna:2015paa, Luna:2016due, White:2016jzc, Chiodaroli:2016jqw,  Goldberger:2016iau, Luna:2016hge, Goldberger:2017frp, Johansson:2017bfl, Johansson:2017srf,  Adamo:2017nia} of the gravity = gauge $\times$ gauge paradigm.




 \section{Strongly Coupled Yang-Mills and  $(2, 0)$ Theories}\label{20}


  The  free $(2,0)$  theory is described by the  $(2,0)$ tensor multiplet consisting of an Abelian two-form  gauge potential $B_{\mu\nu}$ with self-dual three-form field strength $H=\star H$,   four symplectic Majorana-Weyl spinors $\chi$ and five scalars $\Phi$, transforming respectively as the  $\rep{1, 4}$ and $\rep{5}$ of the the rigid $\Spin(5)\cong\USp(4)$ R-symmetry. The two-form gauge and gauge-for-gauge transforms are given by 
\be\label{btran}
\delta B_{\mu\nu}=2\partial_{[\mu}\lambda_{\nu]}, \quad \delta \lambda_{\nu}=\partial_{\nu}\lambda 
\ee
leaving $15-6+1=10$ off-shell degrees of freedom. 
The equation of motion $d\star H=0$ leaves six on-shell degrees of freedom in the $\rep{(3,1)+(1,3)}$ representation of the spacetime little group $\Sp(1)\times\Sp(1)$. The self-duality condition, which with  the Bianchi identity
$dH=0$ implies the equation of motion, further reduces these to the chiral  $\rep{(3,1)}$ representation. Dimensionally reducing on a circle, $S^1$, with radius $R$ yields the maximally supersymmetric Abelian $D=5, \N=4$ gauge theory, consisting  of a one-form Abelian gauge potential $A_m$, four symplectic Majorana spinors $\psi$ and five scalars $\phi$, with coupling constant $g^{2}\propto R$ and the same $\USp(4)$ R-symmetry.

Going beyond the free theories it has been conjectured \cite{Rozali:1997cb, Berkooz:1997cq} that the strong coupling limit of  $D=5$, $\N=4$ Yang-Mills theory is given by an \emph{interacting} $(2, 0)$ theory  compactified on $S^1$ with $g_{YM}^{2}\propto R$. Crucial to this picture is the existence of 1/2-supersymmetric instantonic 0-branes   in the $D=5, \N=4$ Yang-Mills theory, which preserve the full $\USp(4)$ R-symmetry. They have mass $\propto |n|/g_{YM}^{2}$, where $n$ is the instanton number, so that they become light in the strong coupling limit and can be  matched to the  Kaluza-Klein modes  of the $(2,0)$ theory compactified on $S^1$, which have mass $\propto n/R$ \cite{Rozali:1997cb}. 

 \section{Strongly Coupled Gravity and the $(4, 0)$ Theory}\label{40}

Maximally supersymmetric $D=5, \N=8$ supergravity has  $\USp(8)$ R-symmetry and  an exceptional non-compact global $E_{6(6)}(\R)$ symmetry \cite{Cremmer:1980gs} that is broken by quantum effects to the discrete subgroup $E_{6(6)}(\Z)$,  corresponding to the U-duality group of M-theory compactified on $T^6$ \cite{Hull:1994ys}. Its massless fields 
include 27 one-form Abelian gauge potentials $A_{m}$,
 transforming  in the fundamental $\rep{27}$  of $E_{6(6)}$.
Hull \cite{Hull:2000zn, Hull:2000ih, Hull:2000rr} considered a large $l_5$ limit  under the assumption that the $E_{6(6)}$ symmetry is preserved and all supersymmetric states are protected.  Decomposing the $\N=8$ multiplet with respect to an $\N=4$ subalgebra, we obtain five $\N=4$ Abelian gauge multiplets with coupling constant $g^{2}=l_5$,  each of which therefore lifts to an Abelian $(2,0)$ theory as $l_5\rightarrow\infty$, where $g^{2}=l_5$ is identified with $R$ as before. 
If the $E_{6(6)}$ symmetry is to be preserved it follows that \emph{all} 27 one-forms must lift to two-forms. Hence, if all supersymmetries survive the entire $\mathcal{N}=8$ supergravity multiplet must lift to a $D=6$ theory, where $l_5$  is identified with $R$ such that the $l_5\rightarrow\infty$ limit is conformal.  We therefore require a  superconformal gravitational theory in $D=6$ dimensions, consistent with a global $E_{6(6)}$ symmetry, that yields $D=5, \N=8$ supergravity when compactified  on a circle.  According to Nahm's classification there is a unique candidate satisfying these criteria, the $(4,0)$ theory.

As described in \cite{Hull:2000zn, Hull:2000ih, Hull:2000rr} the free $(4, 0)$ theory consists of  eight two-form ``gravitini'', $\Psi_{\mu\nu}$, 27 Abelian self-dual two-forms, $B_{\mu\nu}$, 48 symplectic Majorana-Weyl spinors, $\lambda$, and 42 scalars, $\Phi$,  transforming respectively as the  $\rep{8, 27, 48}$ and $\rep{42}$ of the   $\USp(8)$ R-symmetry.  Finally, rather than a graviton there is a  rank four tensor, 
\be\label{riemman}
G_{\mu\nu\rho\sigma}=G_{[\mu\nu][\rho\sigma]}=G_{[\rho\sigma][\mu\nu]}, \quad G_{[\mu\nu\rho]\sigma}=0,
\ee which might be thought of as a ``gravi-gerbe'' field \cite{Mason:2011nw, Mason:2012va}. It has a  rank six field strength,
\be
R_{\mu\nu\rho\sigma\tau\lambda}=9\partial_{[\mu}G_{\nu\rho][\sigma\tau, \lambda]}=R_{\sigma\tau\lambda\mu\nu\rho},
\ee
satisfying the first and second Bianchi identities,
\be
R_{[\mu\nu\rho\sigma]\tau\lambda}=\partial_{[\kappa}R_{\mu\nu\rho]\sigma\tau\lambda}=0.
\ee
It is invariant under the  gauge transformations,  
\be\label{Ggauge}
\begin{split}
\delta G_{\mu\nu\rho\sigma} &= \partial_{[\mu}\xi_{\nu]\rho\sigma}+\partial_{[\rho}\xi_{\sigma]\mu\nu}-2\partial_{[\mu}\xi_{\nu\rho\sigma]}\\
&= \partial_{[\mu}\zeta_{\nu]\rho\sigma}+\partial_{[\rho}\zeta_{\sigma]\mu\nu}
\end{split}
\ee
where $\quad \xi_{\rho\mu\nu}=\xi_{\rho[\mu\nu]}$ and $\zeta_{\nu\rho\sigma}:=\xi_{\rho\mu\nu}-\xi_{[\rho\mu\nu]}$.
The natural free  field equation,
$
R^{\mu}{}_{\nu\rho\mu\tau\lambda}=0$, 
 describes ten on-shell degrees of freedom in the $\rep{(5, 1)+(1, 5)}$. This is reduced to the chiral $\rep{(5, 1)}$ representation by the  self-duality relation $R=\star R=R\star$.
 It was shown in  \cite{Hull:2000ih} that the  free $(4,0)$ theory compactified on a circle yields linearised $D=5, \N=8$ supergravity. The $(4, 0)$ theory is   gravitational, but does \emph{not} contain a graviton.
 
 As for the $(2,0)$ theory, it is not possible to construct a conventional set of local covariant interactions,  making the non-linear theory difficult to probe. 
 Nonetheless, an analysis of the BPS spectrum  analogous to that of the $(2, 0)$ theory suggests that the identification of the strong  coupling limit of $D=5, \N=8$ supergravity as the full interacting $(4,0)$ theory compactified on $S^1$ is in principle consistent \cite{Hull:2000zn}. In particular, $D=5, \N=8$ supergravity admits 1/2-supersymmetric  gravitational instantonic  solutions, which preserve the $E_{6(6)}$ symmetry \cite{Hull:1997kt}. In analogy with the instantons appearing in the $(2,0)$ story, these are the uplift of Euclidean $D=4$  self-dual gravitational instantons \cite{Gibbons:1979zt, Gibbons:1984hy, Hull:1997kt}, which can be interpreted as 0-branes  \cite{Hull:1997kt}. They carry mass $\propto |n|/l_5$ and so become light in the $l_5\rightarrow \infty$ limit. The analysis of \cite{Witten:1995ex, Rozali:1997cb} indicates that these solutions may be regarded as the Kaluza-Klein modes of a $D=6$ theory on a circle of radius $R\propto l_5$ \cite{Hull:2000zn}. This proposal still requires many checks, but encouragingly, these 1/2-supersymmetric states sit in massive $(4,0)$ multiplets that have  precisely the correct content to have originated from an $S^1$ compactification of the $D=6, (4,0)$ theory: 27 massive self-dual two-forms and 42 massive scalars. A more detailed analysis of the $D=5, \N=8$ and $D=6, \N=(4,0)$ supersymmetric multiplets paints a compelling picture. In particular, the 27 self-dual two-forms in $D=6$ couple to self-dual supersymmetric strings, which yield the  required $D=5$ charged 0-branes and 1-branes transforming in the $\rep{27}$ and $\rep{27}'$ of the global $E_{6(6)}$. 

\section{The $(2, 0)$ Theory Squared}

In direct analogy with \eqref{def} we apply the  product  to  a pair of self-dual two-forms  belonging to  left and right Abelian $(2,0)$ tensor multiplets, 
\be\label{G}
  \mathcal{G}_{\mu\nu\rho\sigma}:=  B_{\mu\nu} \circ \tilde{B}_{\rho\sigma}.
\ee
Adopting  this dictionary  we recover precisely the free $(4, 0)$ theory. In particular, the generalised gauge  transformations of the gravi-gerbe field \eqref{riemman}  are generated by the local symmetries  of the left and right $(2,0)$ factors. Since the supercharges of the left and right  theories  generate the supersymmetries of their product \cite{Anastasiou:2014qba}, the remaining fields of the $(4, 0)$ multiplet and their transformations  then follow essentially automatically.

The  field $\mathcal{G}$ has $15\times 15=225$ components, reduced to $10\times 10=100$ off-shell degrees of freedom by the generalised gauge transformations generated by \eqref{btran}. Explicitly, using $\partial (f\circ g) =\partial f\circ g=f\circ \partial g$ we obtain,
\be\label{Gtrans}
\begin{split}
\delta \mathcal{G}_{\mu\nu\rho\sigma} &= \delta B_{\mu\nu} \circ \tilde{B}_{\rho\sigma}+ B_{\mu\nu} \circ \delta \tilde{B}_{\rho\sigma}\\
&= 2\partial_{[\mu}C_{\nu]} \circ \tilde{B}_{\rho\sigma}+ B_{\mu\nu} \circ 2\partial_{[\rho}\tilde{C}_{\sigma]}\\
&= 2\partial_{[\mu}C^{\io}_{\nu]\rho\sigma} + 2 \partial_{[\rho}{C}^{\oi}_{\sigma]\mu\nu},
\end{split}
\ee
where $\delta$ is the   Becchi-Rouet-Stora-Tyutin (BRST) transformation corresponding to \eqref{btran} and we have introduced the ghost field  dictionary,
\be\label{paras}
C^{\io}_{\nu\rho\sigma}=C_{\nu} \circ \tilde{B}_{\rho\sigma},\qquad {C}^{\oi}_{\sigma\mu\nu}= B_{\mu\nu} \circ \tilde{C}_{\sigma}.
\ee
Here the superscripts $(x\tilde{x})$ denote the ghost numbers of the  left/right factors, which are additive so that the  ghost number of $C^{\scriptsize{(x\tilde{x})}}$ is $x+\tilde{x}$. The  ghosts $C^{\io}_{\nu\rho\sigma}, C^{\oi}_{\nu\rho\sigma}$ have  $6\times 15+6\times 15= 180$  components. However the left/right 2-form ghost-for-ghost transformations, $\delta C_{\nu}=\partial_{\nu}C$, generate gravi-gerbe ghost-for-ghost transformations. Using    $\delta (f^{\scriptsize{(x)}}\circ g^{\scriptsize{(\tilde{x})}})= \delta f^{\scriptsize{(x)}}\circ g^{\scriptsize{(\tilde{x})}} +(-1)^{x} f^{\scriptsize{(x)}}\circ \delta g^{\scriptsize{(\tilde{x})}}$  the full set  of BRST variations and ghost fields can  be systematically determined by repeatedly varying the field  \eqref{G} and ghost \eqref{paras} dictionaries. This procedure yields, 
\begin{subequations}\label{eq:gaugegauge}
\begin{eqnarray}
\delta C^{\io}_{\nu\rho\sigma} &=& \partial_{\nu} C^{\jo}_{\rho\sigma}-2\partial_{[\rho}C^{\ii}_{|\nu|\sigma]} \\
 \delta C^{\oi}_{\nu\rho\sigma} &=& \partial_{\nu} C^{\oj}_{\rho\sigma}+2\partial_{[\rho}C^{\ii}_{\sigma]\nu} \\
   \delta C^{\ii}_{\rho\sigma} &=&  \partial_{\rho} C^{\ji}_{\sigma}-\partial_{\sigma}C^{\ij}_{\rho} \\
 \delta C^{\jo}_{\rho\sigma} &=& 2\partial_{[\rho} C^{\ji}_{\sigma]} \\
 \delta C^{\oj}_{\rho\sigma} &=&2 \partial_{[\rho} C^{\ij}_{\sigma]} \\
 \delta C^{\ji}_{\rho} &=&  \partial_{\rho} C^{\jj}\\
  \delta C^{\ij}_{\rho} &=&  \partial_{\rho} C^{\jj}
\end{eqnarray}
\end{subequations}
where we have introduced the dictionary for the ghost-for-ghost fields,
\begin{gather}\label{ghostghost}\nonumber
 C^{\jo}_{\rho\sigma} = C\circ \tilde{B}_{\rho\sigma}, \quad  C^{\ii}_{\rho\sigma} = C_{\rho} \circ \tilde{C}_\sigma,
 \quad  C^{\oj}_{\rho\sigma} =  B_{\rho\sigma} \circ \tilde{C};\\\label{ghostghost}
   C^{\ji}_{\rho} = C\circ \tilde{C}_{\rho}, \quad\quad
  C^{\ij}_{\rho} =  C_{\rho} \circ \tilde{C};\\\nonumber
  C^{\jj}=C\circ\tilde{C}.
\end{gather}
The complete set of ghost fields removes a total of $125=(90 +90)  -(15 +15 +36)  +(6 +6) -1$ components from $\mathcal{G}$, leaving $100$ off-shell degrees of freedom as expected. That the full set of generalised gauge transformations  is generated  directly by the left/right factors  is a nice feature of the construction.

Let us now define the irreducible $\text{GL}(6, \R)$ representations,
\begin{subequations} \label{3reps}
\begin{align}\label{grav}
G_{\mu\nu\rho\sigma} &= \frac{1}{2}\left(\mathcal{G}_{\mu\nu\rho\sigma}+\mathcal{G}_{\rho\sigma\mu\nu}\right) - \mathcal{G}_{[\mu\nu\rho\sigma]},\\\label{scalar}
\Phi_{\mu\nu\rho\sigma} &=\mathcal{G}_{[\mu\nu\rho\sigma]},\\\label{2form}
\mathcal{B}_{\mu\nu\rho\sigma} &= \frac{1}{2}\left(\mathcal{G}_{\mu\nu\rho\sigma}-\mathcal{G}_{\rho\sigma\mu\nu}\right), 
\end{align}
\end{subequations}
which transform as the $\rep{1}+\rep{20}+\rep{84}$, $\rep{15}$ and $\rep{15}+\rep{45}+\overline{\rep{45}}$ of $\Spin(1,5)$, respectively. 

First, $G_{\mu\nu\rho\sigma}$ has the symmetries of \eqref{riemman} and, directly  from \eqref{Gtrans},  the generalised gauge transformations given in \eqref{Ggauge}, where we have identified the ghost field,
\be
\xi_{\nu\rho\sigma} := C^{\io}_{\nu\rho\sigma}  + C^{\oi}_{\nu\rho\sigma}.
\ee
 Hence, it is naturally identified with the gravi-gerbe field \eqref{riemman} of the $(4,0)$ multiplet.  Note, $G$ has a total of $50=105-70+15$ off-shell degrees of freedom sitting in the $\rep{1+14+35}$ of $\Spin(5)$. This follows directly from the generalised ghost and ghost-for-ghost transformations generated by \eqref{eq:gaugegauge} through the dictionary \eqref{ghostghost}, 
\be
\delta \zeta_{\nu\rho\sigma} =\partial_{\nu} \zeta_{\rho\sigma} + \partial_{[\sigma}\zeta_{\rho]\mu},\qquad \delta \zeta_{\rho\sigma} =0,
\ee
where $\zeta_{\rho\sigma}:=3(\xi_{\rho\sigma}-C^{\ii}_{[\rho\sigma]})/4$ and $2\xi_{\rho\sigma}:=C^{\jo}_{\rho\sigma}+C^{\oj}_{\rho\sigma}$.

Similarly, it is straightforward to show that the left/right two-form gauge symmetries imply that $\Phi_{\mu\nu\rho\sigma}$ has  four-form gauge transformations given by,
\begin{subequations} 
\begin{gather}
\delta \Phi_{\mu\nu\rho\sigma}=4\partial_{[\mu}{\Lambda}_{\nu\rho\sigma]},  \quad\delta {\Lambda}_{\nu\rho\sigma}=3\partial_{[\nu}{\Lambda}_{\rho\sigma]},\\   \delta {\Lambda}_{\rho\sigma}=2\partial_{[\rho}{\Lambda}_{\sigma]}, \quad \delta {\Lambda}_{\sigma}=\partial_{\sigma}{\Lambda}
\end{gather}
\end{subequations} 
where ${\Lambda}_{\nu\rho\sigma}=\xi_{[\nu\rho\sigma]}$,  ${\Lambda}_{\rho\sigma}=\xi_{[\rho\sigma]}+2C_{[\rho\sigma]}^{\ii}$,  ${\Lambda}_{\sigma}=3(C_{\sigma}^{\ji}+C_{\sigma}^{\ij})/2$ and $\Lambda=3C^{\jj}/2$,  leaving $5=15-20+15-6+1$ off-shell degrees of freedom in the $\rep{5}$ of $\Spin(5)$.

Finally, \eqref{2form}
transforms as
\begin{subequations} \label{2formgauge}
\begin{gather}
\delta \mathcal{B}_{\mu\nu\rho\sigma}=\partial_{[\mu}\alpha_{\nu]\rho\sigma}-\partial_{[\rho}\alpha_{\sigma]\mu\nu}, \\
\delta \alpha_{\nu\rho\sigma} =\partial_{\nu} \alpha_{\rho\sigma} -2 \partial_{[\rho}\beta_{\sigma]\nu},\\
\delta \alpha_{\rho\sigma} = 2\partial_{[\rho}\alpha_{\sigma]},\quad\delta \beta_{\sigma\nu} =2 \partial_{(\sigma}\alpha_{\nu)},
\end{gather}
\end{subequations}
where $\alpha_{\nu\rho\sigma}:=C^{\io}_{\nu\rho\sigma}  - C^{\oi}_{\nu\rho\sigma}$, $\alpha_{\rho\sigma}:=C^{\jo}_{\rho\sigma}  - C^{\oj}_{\rho\sigma}$, $\alpha_{\sigma}:=C^{\ji}_{\sigma}  - C^{\ij}_{\sigma}$ and $\beta_{\rho\sigma}:=2C^{\ii}_{(\rho\sigma)}$. This leaves
 $45=105-90+36-6$ off-shell degrees of freedom in the $\rep{10}+\rep{35}$ of  $\Spin(5)$. In total, we have 100 off-shell degrees of freedom in the $\rep{10}\times\rep{10}=\rep{1}_s+\rep{14}_s+\rep{35}'_s +\rep{5}_s+\rep{10}_a+\rep{35}_a$, as expected since each two-form represents a $\rep{10}$ of  $\Spin(5)$. While \eqref{grav} and \eqref{scalar} are immediately recognisable as the off-shell potentials for the  gravi-gerbe \eqref{riemman} and a scalar field (in its dual form), respectively,  \eqref{2form} is perhaps less familiar. It describes the same on-shell degrees of freedom as a self-dual two-form, as is most easily seen by going to physical gauge \footnote{Since the gauge transformations and equations of motions of the factors imply those of the product, one can go to physical gauge in the factors first from which the physical gauge in the product follows immediately.} using the gauge transformations given in  \eqref{2formgauge}. In this case we have $\mathcal{B}_{ijkl}=\mathcal{B}_{[ij][kl]}=-\mathcal{B}_{klij}$, $i,j=1,\ldots, 4$, where the self-duality relations $\mathcal{B}=\star \mathcal{B} =\mathcal{B}\star$ (which follow directly from the left and right self-duality relations in physical gauge $B_{ij} = \star B_{ij}$) leave three independent degrees of freedom in the $\rep{(3,1)}$ of $\Sp(1)\times\Sp(1)$.
 
 Applying global supersymmetries to the factors the rest of the $(4, 0)$ multiplet follows. For example, the eight two-form gravitini $\Psi_{\mu\nu}$ are identified with the eight products, $\chi\circ \tilde{B}_{\mu\nu}$ and $B_{\mu\nu}\circ \tilde{\chi}$ \footnote{We are ignoring various subtleties here. In particular, the $\Gamma$-trace part should be identified with eight of physical symplectic Majorana-Weyl spinors. The $\Gamma$-trace part of $\Psi_{\mu\nu}$ is then re-introduced with a contribution from the left (anti)ghost $\times$ right (anti)ghost sector.}. The super-BRST variation $\delta \Psi_{\mu\nu} = 2\partial_{[\mu} \eta_{\nu]}$ is generated by the left/right two-form  transformations, where the \emph{bosonic}  spinor-vector ghosts $\eta_{\nu}$ are identified with  $\chi\circ \tilde{C}_{\nu}$ and $C_{\nu}\circ \tilde{\chi}$. The complete details will be presented elsewhere. 
 
 Before concluding we note, briefly, that by going first to physical gauge  the equations of motion, Bianchi identities and self-dualities relations for the free $(4, 0 )$ theory follow straightforwardly from those of the $(2, 0)$ factors. Recall, the on-shell degrees of freedom of a self-dual two-form  are given by a symmetric bi-spinor $B_{AB}$, $A,B=1,2$,   in the $\rep{(3,1)}$ of $\Sp(1)\times\Sp(1)$, where $\Box B_{AB}=0$. Hence, for example, the symmetrized product $G_{(ABCD)}=B_{(AB}\circ \tilde{B}_{CD)}$ yields  the $\rep{(5,1)}$ representation satisfying $ \Box G_{(ABCD)}=0$,   which corresponds to the gravi-gerbe field \eqref{riemman} in physical gauge, 
 \be
 G_{ijkl}=G_{[ij][kl]}=G_{klij}, \quad G_{[ijk]l}=0,
 \ee
  where $G_{ijkl}=\star G_{ijkl} = G\star_{ijkl}$ \cite{Hull:2000zn}.

\section{Conclusions}

We have shown that the linear $(4,0)$ theory and its local symmetries follow from the square of Abelian $(2, 0)$ theories. This leaves a number of directions for future work. Perhaps most obvious is the need to understand the $(4,0)$ theory beyond the linear approximation. A natural setting for such a question is higher gauge theory \cite{Baez:2010ya}. For example, a number of higher gauge  $(2,0)$ models were  developed in  \cite{Saemann:2012uq, Saemann:2013pca, Jurco:2014mva, Jurco:2016qwv} using superconformal twistors. However,  the $(4,0)$ theory will require new structures,   gravitational analogs  of the $(2,0)$ models, and it is not a priori clear how to proceed. Here, however,  we have an extra input to guide our considerations: the $(4, 0)$ higher gauge theory will be \emph{required} to be consistent with the square of the $(2,0)$ theory. 

Irrespective, we can still test $(4,0)=(2, 0)\times (2, 0)$ by considering its compatification, in the first instance, on a circle. Besides testing the expected amplitude relations \cite{Czech:2011dk}, we anticipate a matching  of classical solutions, at least in a weak-field approximation,  using  the methodology developed in \cite{Cardoso:2016ngt, Cardoso:2016amd}. In particular, it is natural to  expect that the 1/2-supersymmetric  gravitational instantonic  solutions of $D=5, \N=8$ supergravity, which must be identified with Kaluza-Klein modes of the would-be $(4, 0)$ theory,  are related to the ``square'' of the 1/2-supersymmetric instantonic 0-branes in the $D=5, \N=4$ Yang-Mills theory, which are the Kaluza-Klein modes of the $(2, 0)$ factors. 

We conclude with some rather speculative comments regarding the strong/weak gravitational S-duality suggested by the  $(4, 0)$ theory \cite{Hull:2000zn, Hull:2000ih, Hull:2000rr}. First, note that the generalised gauge invariant curvature, self-duality relations  and Bianchi identities for $\mathcal{G}$ follow directly from those of $B_{\mu\nu}$ and $\tilde{B}_{\rho\sigma}$. In particular, the generalised gauge invariant curvature is the product of the left and right   three-form curvatures,  
\be\label{redfs}
\mathcal{R}_{\mu\nu\rho\sigma\tau\lambda}= 9\partial_{[\mu}\mathcal{G}_{\nu\rho][\tau\lambda, \sigma]}=H_{\mu\nu\rho}\circ \tilde{H}_{\sigma\tau\lambda}.
\ee 
It then follows immediately that the left/right two-form self-duality conditions, $H=\star H, \tilde{H}=\star \tilde{H}$,  and Bianchi identities, $dH=d\tilde{H}=0$ imply  the self-duality relations,
$
\mathcal{R}=\star \mathcal{R}=\mathcal{R}\star$, and the  Bianchi identities, $\partial_{[\mu}\mathcal{R}_{\nu\rho\sigma]\tau\lambda\kappa}=\partial_{[\kappa}\mathcal{R}_{|\mu\nu\rho|\sigma\tau\lambda]}=0$, respectively. Now, recall that a $D=6$ Abelian two-form with self-dual field strength, $H=\star H$,  compactified on $T^2$ yields  an $\SL(2, \Z)$ doublet of $D=4$ one-forms $A^i$,  $i=1,2$, which are related through 
$
 F^i = \star F^j\varepsilon_{jk}\gamma^{ki}, 
$
where  $\gamma^{ki}$ is the constant metric on $T^2$. Since the  gravi-gerbe field-strength originates from   $H\circ \tilde{H}$, feeding this observation into the $(2,0)\times (2,0)$  construction  we  anticipate an $\SL(2, \Z)$ triplet of $D=4$ linearised Riemann tensors, 
\be \mathcal{R}^{(ij)}\sim F^{(i}\circ\tilde{F}^{j)}, \ee
obeying the duality constraint $\mathcal{R}^{(ij)}= \star \mathcal{R}^{(kj)}\varepsilon_{jk}\gamma^{ki}$. This is indeed the case: the  free $(4,0)$ theory compactified on $ T^2$ yields  linear $\N=8$ supergravity, with an $\SL(2, \Z)$ symmetry acting on a triplet of duality related  gravitational field-strengths \cite{Hull:2000zn, Hull:2000ih, Hull:2000rr}. Here it is shown to be  the ``square'' of the familiar $\SL(2, \Z)$ of the Abelian $(2,0)$ multiplet compactified on $T^2$. Of course, this symmetry is broken by  interactions.   This is not, however, necessarily an argument against its existence; it simply tells us that  it is not a symmetry of classical $\N=8$ supergravity, just as S-duality is not a symmetry of classical $\N=4$ super Yang-Mills theory.  While this picture is suggestive, it is highly speculative and will depend crucially on the non-linear structure of the complete (4, 0) theory. Clearly it may fail to materialise and a strong degree of scepticism is advised, but the lessons in gauge theory and gravity learnt on the journey will regardless return many insights. Even more speculatively,  if the $(4,0)$ theory on $M^6= X\times C$, where $C$ is a punctured Riemann surface,  admits  quantities that are protected as we  vary the size of $X$ or $C$,   then one might expect a gravitational analog, or  square, of the AGT correspondence.

We are grateful to Michael J.~Duff for illuminating conversations. The work of LB is supported by a Schr\"odinger Fellowship.


%

\end{document}